\documentclass[conference]{IEEEtran}
\IEEEoverridecommandlockouts
\usepackage{cite}
\usepackage{amsmath,amssymb,amsfonts}

\usepackage{graphicx}
\usepackage{textcomp}
\usepackage{xcolor}

\usepackage{algorithm}
\usepackage{algorithmic}
\usepackage{array}
\usepackage{nomencl}
\usepackage{eurosym}
\usepackage{multirow}
\usepackage{eurosym}

\usepackage{colortbl}

\usepackage{amsmath} % Required for the align environment
\usepackage{hyperref} % Optional for better referencing

\def\BibTeX{{\rm B\kern-.05em{\sc i\kern-.025em b}\kern-.08em
    T\kern-.1667em\lower.7ex\hbox{E}\kern-.125emX}}
\begin{document}

\title{Balancing service provision by EV aggregator in different TSO-DSO coordination schemes\\
% Reserve flexibility provision by EV aggregator in different TSO-DSO coordination schemes\\
% A comparative study of different TSO-DSO coordinations in using balancing services from EV aggregators
% Participation of Electric vehicle Aggregator in different TSO-DSO coordination schemes\\

\thanks{The authors would like to acknowledge the financial support for this work from the Netherlands Organization for Scientific Research (NWO) funded DEMOSES project.}
}
\author{\IEEEauthorblockN{Hang Nguyen}
\IEEEauthorblockA{\textit{Electrical Energy Systems Group} \\
\textit{Eindhoven University of Technology}\\
Eindhoven, The Netherlands \\
t.h.nguyen@tue.nl}
\and
\IEEEauthorblockN{Phuong Nguyen}
\IEEEauthorblockA{\textit{Electrical Energy Systems Group} \\
\textit{Eindhoven University of Technology}\\
Eindhoven, The Netherlands \\
P.Nguyen.Hong@tue.nl}
\and
\IEEEauthorblockN{Koen Kok}
\IEEEauthorblockA{\textit{Electrical Energy Systems Group} \\
\textit{Eindhoven University of Technology}\\
Eindhoven, The Netherlands \\
j.k.kok@tue.nl}
}

\maketitle

\begin{abstract}
The increasing penetration of Distributed Energy Resources (DERs) in the distribution system has led to the emergence of a new market actor - the aggregator. 
Aggregator serves as a facilitator, enabling flexibility asset owners to access various markets. Among them, electric vehicle (EV) aggregators are gaining more attention due to their expanding use and potential to provide services across multiple markets, particularly in the reserve market.
% the transmission system operator (TSO) uses the DER flexibility under conditions of location and specific device information scarcity. 
Currently, the transmission system operator (TSO) indirectly utilizes the resources under the management of the distribution system operators (DSO), which can negatively impact the distribution grid. 
Conversely, adjustments from DSOs can impact service provision to TSO due to the shortage of TSO usage information. These factors highlight the importance of evaluating the service provision from aggregators under different TSO-DSO coordination schemes.
% Some studies focus on the DSO-managed coordination scheme while the TSO-DSO hybrid-managed offers distinct advantages and aligns well with the prevailing situation in many countries. Therefore, 
This paper focuses on the provision of flexibility from electric vehicles (EVs) aggregators for balancing service in the TSO-DSO hybrid-managed scheme and compares it to the DSO-managed coordination schemes.
The behavior of aggregators reacting to price fluctuations and TSO requests under different coordination schemes and simulation scenarios is thoroughly assessed. Additionally, their impact on the grid is analyzed through the DSO’s congestion management process and validated using data from a real part of the Dutch distribution network. 
Results find that the hybrid-managed coordination scheme gives more benefit to the aggregator than the DSO-managed scheme, and the price paid for the BRP for the deviation has a large impact on the aggregator's behavior.
% The results demonstrate that effective coordination between TSO-DSO in leveraging DER flexibility is essential. 

\end{abstract}


\section{Introduction}

% Nowadays, researching the provision of flexibility from aggregators is no longer a novel topic because it has already been implemented in some European countries like France, Finland, Hungary, Estonia, Denmark, Belgium, Romania, etc \cite{17}. 
Along with exploiting the potential of distributed energy resources (DERs), providing flexibility services with these sources attracts the attention of many organizations and research groups and has already been implemented in some European countries like France, Finland, Hungary, Estonia, Denmark, Belgium, Romania, etc \cite{17}.
The role of the aggregator - a market participant responsible for bundling flexibility from multiple small customers and producers into a portfolio and offering the combined capacity in other markets\cite{18} - is becoming increasingly important.
Many research studies focus on optimal biding strategies\cite{3,4,5,6,7,8}, the provision of services to multiple markets (e.g., spot market, reserve,\cite{3,9,10,14,15} congestion management services\cite{7} or balancing portfolios\cite{3}) and the utilization of diverse appliances including electric vehicles (EVs), battery energy storage system (BESS), heat pumps, thermal-controlled load or their combinations.

Recently, EV aggregators have been receiving increasing attention due to their expanding use, driven by environmental factors and government incentives. 
In 2023, the Netherlands was the sixth-largest EV market in Europe with over 300 thousand battery electric vehicles registered and leads in EV infrastructure \cite{EV_Dutch_1}. 
% In September of 2024, a new subsidy scheme to help fund private charging infrastructure for commercial vehicles was announced by the Netherlands Enterprise Agency prove their focus on EV development in the upcoming years \cite{EV_Dutch_2}.
Research on the Dutch situation shows the potential of EV aggregators to provide service in the reserve market\cite{10}.
EV fleets also demonstrate potential to provide flexibility to the internal balancing portfolio\cite{3}, and congestion management \cite{7}. 
% Results from \cite{5} show that EV is the main source of flexibility and revenue. 
% Via a case study in Austria, Christoph \cite{7} provides important insights that using flexible resources from EVs fleets as a redispatch measure leads to reduced curtailment of renewable energy while less additional thermal power plant usage is needed.
% Miniti et al \cite{8} propose hybrid stochastic programming for optimizing the bidding strategy of EV aggregator, which applied a robust approach in the first stage instead of scenario generation, leading to reduced computation time while bringing a better estimate of the actual daily energy cost than the two-stage stochastic approach. 
The above studies show that EV aggregators have great prospective to provide flexibility services and bring benefits not only for aggregators but also for prosumers and system operators. 
% However, most of them consider the EVs under the virtual EV while ignoring the state of charge limit of each EV. 
% using resources under the management of the distribution system operator (DSO) to provide service to the transmission system operator (TSO).
Therefore, the EV aggregator is investigated in this research to consider its behavior on different market price fluctuations.
% interaction with different transmission system operator (TSO) and distribution system operator (DSO) coordination.

Moreover, previous research has primarily focused on the aggregator business model when providing service to the transmission system operator (TSO), with limited attention given to the impact on the grid under the management of the distribution system operator (DSO). While an effective approach to leveraging DERs flexibility is enhancing TSO-DSO coordination \cite{16}. In recent years, this coordination has become a prominent research trend, attracting significant interest from the academic community.
% This coordination primarily involves the establishment of a common data platform, the sharing of metering data, and network planning to enhance five key aspects:
% (1) Voltage regulation, (2) Reactive power management, (3) Operational cost optimization, (4) Operational planning and (5) Congestion management \cite{2}. 
% Various coordination schemes have been proposed to identify optimal solutions for these purposes via multiple European projects like SmartNet \cite{smartnet}, CoordiNet, Interrface \cite{coordinet}, InteGrid \cite{intergrid} and national projects such as Gopacs \cite{gopacs}, PicoFlex \cite{picoflex} or Soteria \cite{soteria}. 
Arthur et al \cite{1} classify the existing coordination into three main schemes (1) TSO-managed; (2) DSO-managed; and (3) TSO-DSO Hybrid. 
In practice, the DER service provider contracts directly with the TSO in the first model. TSO is the single buyer and is prioritized in accessing DER flexibility and performs economic dispatch of DERs, while the DSO is restricted from accessing resources connected to their network. With less interaction, the implementation of this scheme is straightforward in a centralized manner. However, the distribution network constraints are not considered due to the lack of information. This scheme also poses great computational and modeling challenges to TSO, and a large amount of operational data needs to be shared.
Conversely, the DSO-managed model grants the DSO a more prominent role and greater responsibility in optimizing the use of flexibility. The DSO is given priority to aggregate and employ the DER flexibility for managing congestion in their grids, then send the remaining flexibility to the central market. With increasing RES, it becomes more difficult to implement in a decentralized way via DSOs. Moreover, this scheme has the potential DSO conflicts of interest and transfers the computational challenges to DSO.
Among these 3 schemes, the Hybrid TSO-DSO managed model appears to be the most balanced option as TSO and DSO manage their own market and allow TSO to use service with DSO's consent. The aggregator sends their bid to TSO, then TSO asks DSO to validate the bid. This coordination aligns well with the prevailing circumstances in many countries and offers distinct advantages over others by empowering DSOs role and fostering greater social welfare. Therefore, the DSO-TSO hybrid-managed model is one of the schemes selected for investigation in this research.

% why agg vs TSO-DSO
Considering the provision of DERs flexibility from the aggregator within the TSO-DSO coordination schemes offers a broader perspective by highlighting the risk of violating grid constraints when procuring flexibility for the TSO.  Therefore, this research constructs a comprehensive model in which an EV aggregator provides multiple services under different TSO-DSO coordination schemes.
% However, most of studies focus on 
To the best of the authors's knowledge, this topic has not been widely explored in the literature. This paper shares some similarities with the 3 papers \cite{11,12,13} in using flexibility from the distribution grid via the aggregator while accounting for the TSO-DSO coordination. However, unlike \cite{11}, this paper models the details of the EV aggregator, improves the DSO model with linearized ACOPF and compares the flexibility provision in 2 different coordination schemes. In \cite{12}, the authors show that 16\% and nearly 52\% of the operational cost is reduced for TSO and DSO, respectively. 
While the aggregator experienced a profit loss of approximately 54\% under the DSO-managed model compared to the case without coordination, the study did not explore how the outcomes would differ if the TSO utilized flexibility from the aggregator in the balancing market.
Compared with \cite{13}, this paper models the EV aggregator that considers individual EV to ensure EV energy within the acceptable limit range while providing vehicle-to-grid service and considering the TSO-DSO hybrid-managed model that advances over the DSO-managed coordination scheme. 
% The authors in \cite{15} consider the provision of DER flexibility with grid constraints by adopting the concept of transactive energy. However, this framework considers the interaction between aggregator and DSO while not considering the interaction between TSO and DSO.

The contributions of this work are as follows:
\begin{itemize}
\item Modeling the interaction between TSO, DSO and EV aggregator in the provision of balancing service (aFRR), while considering the constraint in the distribution grid.
\item Considering the service provision in different TSO-DSO coordination schemes and validating the mechanism to one part of the distribution network of the Netherlands.
\item Analysing the behavior of multiple EV aggregators in price fluctuations and seasonal changes.
% response to day-ahead price and imbalance price fluctuation and BRP contracted price.
\end{itemize}
The remainder of the paper is organized into four main parts. Firstly, Section II introduces the interaction of aggregator, TSO and DSO in different coordination schemes. The mathematical model of EV aggregator, TSO and DSO is presented in Section III. Section IV assesses and discusses the aggregator behavior and cost analysis via case study and simulation results.  Finally, the conclusions are presented in Section V.

\section{TSO-DSO coordination}
% Introdude the TSO, DSO, aggregator (in physical layer) (thể hiện sự tương tác theo số. trong text giải thích số đó nghĩa là gì.
Coordination between TSOs and DSOs can take place in various stages, including pre-qualification, offering, and real-time activation of services. 
% Traditionally, TSOs and DSOs share measured power flows at the point of connection, forecast data, and relevant data in emergency situations, but this sharing is not a regular occurrence. With the advent of flexible services, they should also share pricing signals, scheduling, and activation controls of their services to ensure transparency, prevent harmful conflicts in service activation, and avoid double counting \cite{19,20,21,22,23}.
In the context of the Dutch balancing service, coordination is primarily limited to the pre-qualification phase, where equipment specifications are reviewed before granting grid connection approval. However, no interaction occurs during the offering and activation stages, which are conducted in real-time or near real-time
% Balancing service is normally provided to TSO from the large generator and demand via a balancing service provider (BSP) \cite{21}. A BSP offers a balancing service to TSO to balance out any unpredicted imbalances in the electricity grid. Two balancing services are used in the Netherlands, automatic frequency restoration reserve (aFRR) and manual frequency restoration reserve (mFRR). aFRR is frequently used to correct the imbalance within 15 minutes, while mFRR is used for larger and longer disturbances. 
% aFRR has 2 types, "contracted bid" and "free-bid". "Contracted-bid" have to be submitted before 14:45 pm of the prior day to ensure that there are always sufficient balancing bids available, and "free-bids" from non-contracted BSPs can be submitted in 2 imbalance settlement period (IPS) before real-time. 
% Currently, BSP also takes the role of aggregator to provide these services by combining resources from scattered customers.
In this study, coordination in the offering process for automatic frequency restoration reserve (aFRR) provision is proposed with two different schemes: the DSO-managed model and the TSO-DSO hybrid-managed model. 
The TSO-managed model places a significant computational burden on the TSO and limits the DSO’s ability to flexibly utilize services from resources connected to its own network. This restriction can lead to reduced efficiency and fairness in the facilitation of DER flexibility services. As a result, this type of coordination is considered beyond the scope of this study.
% Depend on the service they provide, aggregator can be balancing service provider or congestion service provider (CSP).
% (but CSP and market close in the day-before, a long duration with uncertain of price or generation and demand error.
% According to \cite{23}, aggregators must be able to undertake value stacking, it means they can provide multiple services or to multiple flexibility responsible parties.
% From the above discussion,
\subsection{TSO-DSO hybrid-managed coordination scheme}
\begin{figure}[b]
\centerline{\includegraphics[scale=0.12]
{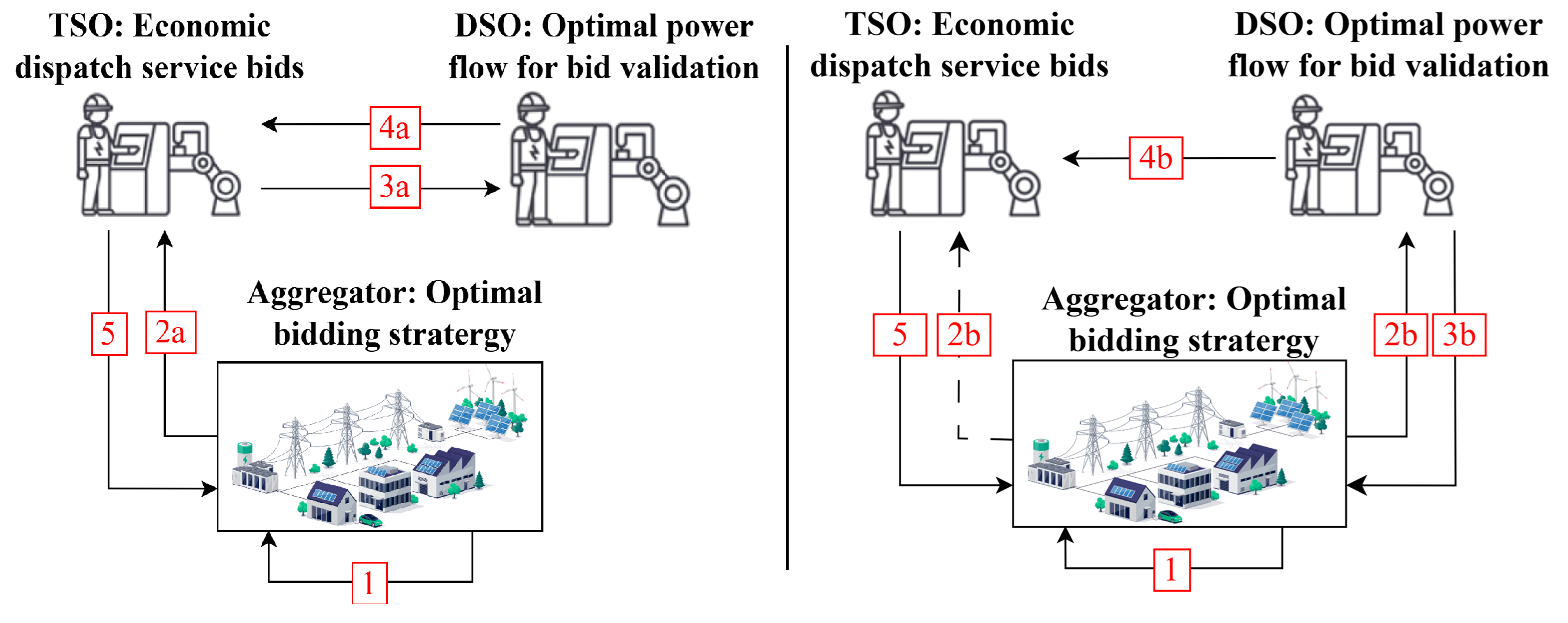}}
\caption{a) TSO-DSO hybrid-managed, and b) DSO-managed coordination schemes.}
\label{fig:fig0_0}
\end{figure}

The first scheme is illustrated in Fig. \ref{fig:fig0_0}a on the left, which consists of 5 main processes. In which,  (1) the aggregator aggregates all the resources from their prosumers, determines the optimal bidding strategy and sends it to the balance responsible party (BRP) before the energy market gate closing time. (2a) The aggregator also defines the portion of flexibility to be offered as balancing services and sends this information to the TSO. (3a) The TSO conducts the economic dispatch and compiles the merit order list (MOL) (a list of offers sorted by price from low to high) for the next two imbalance settlement period (ISP) and forwards it to the DSO. (4a) Upon receiving the MOL, the DSO performs the validation process. This involves eliminating invalid bids/offers by perform power flow analysis to determine the grid congestion, following the traffic light concept described in \cite{soteria}. The green state indicates that no congestion occurs, while the yellow state signals that congestion is happening in their network. DSO updates the upper and lower boundary of the flexibility by using linearized ACOPF and communicates the validated result back to the TSO. (5) Finally, TSO activates the service in real time.

\subsection{DSO-managed coordination scheme}
% \begin{figure}[b]
% \centerline{\includegraphics[scale=0.12]{images/TSO_DSO_aggregator_3.png}}
% \caption{DSO-managed coordination scheme.}
% \label{fig:fig0_1}
% \end{figure}
Fig. \ref{fig:fig0_0}b on the right illustrates the schemes where the DSO is prioritized in utilizing flexibility. The sequence of information exchange is structured into four main processes. (1) The aggregator optimizes its bidding strategy and defines the volume available for upward and downward regulation and (2b) These bids and offers are then submitted to the DSO for validation. Upon validation, DSO sends the volume they use to resolve congestion to the aggregator (3b) and the new boundary of the upward/downward volume that the aggregator can provide to TSO (4b). Based on the new boundaries, TSO constructs the MOL and sends activation requests to the aggregator (5).
% \begin{enumerate}
%  \item Aggregator optimal bidding strategy and define the volume they can bid for upward and downward regulation.
%  \renewcommand{\labelenumi}{2b)}
%  \item Aggregators send the volume they can bid for upward and downward regulation to \textbf{DSO}.
%  \renewcommand{\labelenumi}{3b)}
%  \item DSO sends the new boundary of the upward/downward volume that the aggregator can provide to TSO.
%  \renewcommand{\labelenumi}{4)}
%  \item TSO defines the MOL and sends activation requests to the aggregator in real-time.
% \end{enumerate}

% \begin{algorithm}
% \caption{DSO-managed}\label{euclid}
% \begin{algorithmic}
% \small{
% % \Procedure
% \STATE{\bfseries Input}: flexibility data, activated volume
% \STATE{\bfseries Output}: $MOL$
% \STATE $DSO \gets $ Aggregator sends the boundaries of their flexibility 
% \STATE $state \gets $ DSO check congestion with maximum flexibility value
% \STATE{\bfseries While}{ $state == Yellow$}
%     \STATE \hspace{10pt} $New\_boundary \gets $ DSO update boundary
%     \STATE \hspace{10pt} $New\_state \gets$ DSO check congestion
%     \STATE \hspace{10pt} $state \gets New\_state$
% % \EndWhile\label{euclidendwhile}
% \STATE $TSO \gets $ DSO send new boundaries
% \STATE $MOL \gets $ TSO performs economic dispatch with new boundaries
% % \STATE \textbf{return} $New\_MOL$
% % \EndProcedure
% }
% \end{algorithmic}
% \end{algorithm}
% Different from Algorithm 1, in Algorithm 2, 
The key difference between this coordination scheme with the previous one lies in steps (2b) and (3b), where the aggregator sends its flexibility boundaries to DSO for validation, rather than directly to the TSO, then DSO sends back the volume they use to resolve congestion to the aggregator. DSO updates the grid's generation and demand profiles by incorporating all available flexibility and performs power flow analysis to check the congestion state in the network. Subsequently, the DSO applies linearized ACOPF to determine updated boundaries of flexibility. This process is repeated until no congestion is detected, the DSO transmits new boundaries to the TSO. Finally, TSO defines the MOL based on the new boundaries. Unlike the DSO-managed model in \cite{1}, the TSO will send its dispatch command to the DER aggregator without DSO acting as an intermediate, due to the dispatched bid is assumed unchanged through the DSO.

In the above processes, the aggregator performs its functions the day before, while the TSO and DSO communicate data during the offering of the aFRR services. The service cost and benefit are determined when the market closes. 
Moreover, when providing services to the TSO, the aggregator deviates from the day-ahead schedule submitted to the BRP in the day before \cite{usef_doc}.
% Broker model, step 4 (page 31 in USEF document)
% Depending on the type of aggregator implementation model, the amount of the deviation cost is compensated in different ways. In the broker model, the aggregator has a contract with the BRP to compensate for the deviations they caused, or in the corrected model, the Meta Data Company will correct the meter information from the connection point with the increase/decrease in energy triggered by the aggregator and inform the TSO of the adjustment. Besides, in the central settlement model, an Allocation Responsible Party will based on a pre-defined price formula to define the amount to be paid by the party to which the energy is transferred. In the case of the Netherlands, TenneT will be the party that informs the BRP about the provision of the BSP service to avoid counteraction and the imbalance cost will not be charged to the BRP anymore.\cite{mFRR_doc}.
% This study focuses on the broker model, where 
As a result, the aggregator incurs a deviation fee payable to the BRP to compensate for their imbalance. This fee influences the aggregator's decisions regarding how much flexibility to reserve. In addition to the deviation fee, the TSO's minimum bid size poses another barrier for aggregators. To enable greater DER flexibility, the TSO should consider lowering the minimum bid size and increasing procurement times, thereby helping to mitigate uncertainty stemming from forecast errors \cite{smarten}.
Therefore, this study assumes that 
% (1) the aggregator has perfect knowledge of the EV session and market price, (2) 
TSO will use all resources provided by the aggregator.
% , and (3) Customers agree with the aggregator to control the charging/discharging of their EV. 
\section{Mathematical models}
The mathematical formulations of the TSO, DSO and aggregator models will be discussed in the following subsections.
\subsection{Aggregator model}
The aggregator links prosumers and markets by determining the optimal bidding schedule to purchase electricity from the day-ahead market to supply prosumers and procuring services from prosumers to offer in other markets. The aggregator's objective is to maximize total benefit when providing services. The objective function and the associated constraints, defined over the whole day (96 steps), are presented below:
\begin{equation}
% F^{Agg} = 
max\sum_{t}^{T}\sum_{i}^{N} E_{i,t}^{\uparrow} (\lambda_t^{\uparrow} - \lambda^{\text{brp}}) 
+ E_{i,t}^{\downarrow}  (\lambda_t^{\downarrow} + \lambda^{\text{brp}}) + E_{i,t}^{\text{da}} (\lambda_t^{\text{da}}-\lambda^{\text{p}})\label{eq:agg_obj}\\
\end{equation}
% \end{equation}
The following constraints need to be satisfied:
\begin{align}
% \small{
% \sum_{t\in T} E_{i,t}^{\uparrow} = 0  \quad\forall i, t &\in [T_{\text{d},i},T_{\text{a},i}]) \label{eq:agg_sum_up_trip}\\
% \sum_{t\in T} E_{i,t}^{\downarrow} = 0 \quad\forall i,t & \in [T_{\text{d},i},T_{\text{a},i}]) \label{eq:agg_sum_down_trip}\\
% \sum_{t\in T} E_{i,t}^{\text{da}} = 0 \quad\forall i,t & \in [T_{\text{d},i},T_{\text{a},i}]) \label{eq:agg_sum_da_trip}\\
P^{\uparrow/\downarrow/da}_{i,t} = E^{\uparrow/\downarrow/da}_{i,t}/\Delta T &\quad\forall {t,i} \label{eq:P2E_up}\\ 
% P^{\downarrow}_{i,t} = E^{\downarrow}_{i,t}/\Delta T &\quad\forall {t,i} \label{eq:P2E_down}\\ 
% P^{da}_{i,t} = E^{da}_{i,t}/\Delta T &\quad\forall {t,i} \label{eq:P2E_da}\\
\begin{cases} 
    P^{\uparrow}_{i,t} = 0 &\text{if } \lambda^{\uparrow}_t = 0,\\
    \underline{P_i^{\uparrow}}u_{i,t}\leq P^{\uparrow}_{i,t} \leq \overline{P^{\uparrow}_i}u_{i,t}  &\text{if } \lambda^{\uparrow}_t \neq 0
\end{cases} &\quad \forall t, i \label{eq:Plim_up}\\
\begin{cases}
    P^{\downarrow}_{i,t} = 0 & \text{if } \lambda^{\downarrow}_t = 0,\\
    \underline{P_i^{\downarrow}}v_{i,t} \leq P^{\downarrow}_{i,t} \leq 
    \overline{P^{\downarrow}_i}v_{i,t} & \text{if } \lambda^{\downarrow}_t \neq 0 
\end{cases} &\quad \forall t, i \label{eq:Plim_down}\\
\underline{P^{\downarrow}_i}w_{i,t} \leq P^{da}_{i,t} \leq \overline{P^{\downarrow}_i}w_{i,t} &\quad \forall t, i \label{eq:Plim_da}\\
% \underline{E_i^{\uparrow}} \leq E_{i,t}^{\uparrow} \leq \overline{E_i^{\uparrow}} &\quad \forall i,t \label{eq:agg_up_limit}\\
% \underline{E^{\downarrow}} \leq E_{i,t}^{\downarrow} \leq \overline{E^{\downarrow}} &\quad \forall i,t \label{eq:agg_down_limit}\\
% \underline{E_i^{\downarrow}} \leq E_{i,t}^{\text{da}} \leq \overline{E_i^{\downarrow}} &\quad \forall i,t \label{eq:agg_da_limit}\\
% E^{\uparrow}(t) - E^{\downarrow}(t) - E^{\text{da}}(t) &= E^{\text{trip}} \quad (t \in [T_{\text{arrive}}, T_{\text{max}}]) \label{eq:agg_trip}\\
% no charge, discharge at the same time
u_{i,t} + v_{i,t} + v_{i,t} \leq 1 &\quad\forall {t,i} \label{eq:cond_up}\\
% E^{\uparrow}_{i,t}\times E^{\downarrow}_{i,t} = 0 &\quad\forall {t,i} \label{eq:cond_up}\\
% E^{\uparrow}_{i,t}\times E^{\text{da}}_{i,t} = 0 &\quad\forall {t,i} \label{eq:cond_da}\\
% E^{\downarrow}_{i,t}\times E^{\text{da}}_{i,t} = 0 &\quad\forall {t,i} \label{eq:cond_down}\\
% EV soc
0.2 \leq E^{\text{EV}}_{i,t}/{E^{\text{EV,Size}}_i} \leq 1 &\quad\forall {t,i} \label{eq:agg_soc_limit}\\
% energy by t
E^{\text{EV}}_{i,t} = E^{\text{EV,Size}}_{i} \quad \forall i,t=0/ T_{d,i}/95 &\label{eq:agg_init}\\
% % t depart
% E^{\text{EV}}_{i,t} = E^{\text{EV,Size}}_i \quad\forall i, t=T_{d,i}&(\neq 0) \label{eq:agg_dep}\\
% % t mid night
% E^{\text{EV}}_{i,t} = E^{\text{EV,Size}}_i \quad\forall i, t=95&(\neq T_{\text{a}}) \label{eq:agg_mid}\\
% t at home
E^{\uparrow}_{i,t} + E^{\downarrow}_{i,t} + E^{\text{da}}_{i,t} =E^{\text{EV}}_{i,t-1}-E^{\text{EV}}_{i,t} \quad \forall i, t &\notin [T_{\text{d},i},T_{\text{a},i}] \label{eq:agg_at_home} \\
% t leave home
E^{\text{EV}}_{i,t} = E^{\text{EV}}_{i,t-1} - E^{EV,trip}_i/TL \quad \forall i, t &\in [T_{\text{d},i},T_{\text{a},i}] \label{eq:agg_leave_home}
% \quad (t \in (T_{\text{init}}, & T_{\text{leave}}]) \& (T_{\text{arrive}}, T_{\text{max}}]) \label{eq:agg_at_home}
% } 
\end{align}

% %%%
% 0.2 \leq E^{\text{EV}}_{i,t}/{E^{\text{EV,Size}}_i} \leq 1 &\quad\forall {t,i} \label{eq:agg_soc_limit}\\
% % energy by t
% E^{\text{EV}}_{i,t} = E^{\text{EV,init}}_{i} &\quad \forall i,t=0 \label{eq:agg_init}\\ 
% % t leave
% % E^{\text{EV}}_{i,t} = E^{\text{EV,size}}_{i} &\quad\forall i,t=T_{\text{dep} and t \neq 0 \label{eq:agg_max}\\
% % t at home
% E^{\uparrow}_{i,t} + E^{\downarrow}_{i,t} + E^{\text{da}}_{i,t} =E^{\text{EV}}_{i,t-1}-E^{\text{EV}}_{i,t} &\quad \forall i, t \notin [T_{\text{dep}},T_{\text{arr}}] \label{eq:agg_at_home} \\
% %%%

The volumes of flexibility of each EV for upward and downward and the volume of energy bought from the day-ahead market are $E_i^{\uparrow}$, $E_i^{\downarrow}$, and 
$E_i^{\text{da}}$ accordingly. The set of $\lambda^{\uparrow}$, $\lambda^{\downarrow}$ and $\lambda^{\text{da}}$ are balancing prices of flexibility for regulating upward, downward and the day-ahead price, respectively. $\lambda_{\text{brp}}$ represents the contracted price that the aggregator pays to the BRP for their deviation from the day-ahead schedule. $\lambda_{\text{p}}$ denotes the price at which the aggregator sells electricity to the EV's owner for charging. And $N$ refers to the number of EVs considered, while T indicates the number of time step considered in the next day, which is set to 96 in this study.
% Aggregator considers Eq. \ref{eq:agg_sum_up_trip}, Eq. \ref{eq:agg_sum_down_trip} and Eq. \ref{eq:agg_sum_da_trip} to make sure the EV will not interact with the electricity market when EVs are away from home, it means the total of energy provided for upward or downward regulation or buy from the DA market is zeros after the departure time $(T_{d})$ and before arrival time $(T_{a})$. 
Each EV is assumed to have a single departure session per day and to charge exclusively at home. This implies that there is no interaction with the electricity market after the departure time $(T_{d})$ and before the arrival time $(T_{a})$. 
Eq. \ref{eq:P2E_up} presents the relationship between power $(P_i)$ and energy $(E_i)$ within one ISP, $\Delta T$ is 0.25 hour. Besides, power charge/discharge is limited to the min/max charging rate $(\underline{P_i^{\downarrow}},\overline{P_i^{\downarrow}})$ or min/max discharging rate $(\underline{P_i^{\uparrow}},\overline{P_i^{\uparrow}})$  by Eq. \ref{eq:Plim_up}, \ref{eq:Plim_down} and \ref{eq:Plim_da}. 
Constraint \ref{eq:cond_up} ensures that the aggregator will provide multiple services with the "In time" stacking type. This implies that the EV is neither provides upward nor downward service nor buys from the DA market at the same time step. This exclusivity is ensured through the use of the binary variables $u_{i,t}, v_{i,t}$, and $w_{i,t}$.
Moreover, the state of charge (SOC) of each EV is constrained by Eq. \ref{eq:agg_soc_limit}, ensuring that SOC does not exceed the maximum capacity (100\%) and does not fall below 20\%. This lower bound prevents deeper discharges, which can lead to accelerated wear and reduced lifespan of the battery.
At the beginning of the day, at the leaving moment, and at midnight, EV energy $(E^{\text{EV}}_{i,t})$ is initialized to be fully charged in Eq. \ref{eq:agg_init}. 
Equation \ref{eq:agg_at_home} updates the EV's energy by step when the EV is at home, EV energy at step $t$ is equal to the energy from the previous step, incorporating the change from total energy charge and discharge. When EV leaves home, energy is assumed to decrease gradually by the total energy for the trip ($E_i^{EV,trip}$) divided by the trip length ($TL$) via Eq. \ref{eq:agg_leave_home}.

\subsection{TSO model}
TSO leverages the DER flexibility to minimize the cost of balancing the system for each ISP by the function below:

\begin{align}
% \small{
% Objective function
% F_{t}^{TSO} = 
min\sum_{m\in M} \
E_{m,t}^{a,\uparrow}\lambda_{m}^{a,\uparrow} -\sum_{k\in K} E_{k,t}^{a,\downarrow}\lambda_{k}^{a,\downarrow} &+ E_t^{r,\uparrow}\lambda_{t}^{\uparrow} - E_t^{r,\downarrow}\lambda_{t}^{\downarrow}\label{eq:TSO_obj}
\end{align}
% upward regulation
The TSO has to satisfy the following constraints:
\begin{align}
\sum_{m\in M} E_{m,t}^{a,\uparrow} + E^{r,\uparrow}_t = E_{t}^{reg,\uparrow} &\quad \forall {m,t} \label{eq:sum_equal_activated_volume_up}\\
% downward regulation
\sum_{k\in K} E_{k,t}^{a,\downarrow} + E^{r,\downarrow}_t = E_{t}^{reg,\downarrow} &\quad\forall {k,t} \label{eq:sum_equal_activated_volume_dn}\\
% up, down limit
\underline{E_{m,t}^{a,\uparrow}} \leq E_{m,t}^{a,\uparrow} \leq \overline{E_{m,t}^{a,\uparrow}} &\quad\forall {m,t} \label{eq:up_agg_lim} \\
\underline{E_{k,t}^{a,\downarrow}} \leq E_{k,t}^{a,\downarrow} \leq \overline{E_{k,t}^{a,\downarrow}} &\quad\forall {k,t} \label{eq:dn_agg_lim}
\end{align}
\begin{align}
E_{k,t}^{a,\downarrow} \leq 0 \leq E_{m,t}^{a,\uparrow} &\quad\forall {m,k,t} \label{eq:agg_sign}\\
E_{t}^{r,\downarrow} \leq 0 \leq E_{t}^{r,\uparrow} &\quad\forall {t} \label{eq:reserve_sign}\\
\overline{E_{m,t}^{a,\uparrow}} = \sum_{n\in N} E_{n,t}^{\uparrow} &\quad\forall {m,t} \label{eq:agg_upper_bound}\\
\underline{E_{k,t}^{a,\downarrow}} = \sum_{n\in N} E_{n,t}^{\downarrow} &\quad\forall {k,t} \label{eq:agg_lower_bound}\\
\underline{E_{m,t}^{a,\uparrow}} = \overline{E_{k,t}^{a,\downarrow}} = 0 &\quad\forall {m,k,t} \label{eq:agg_middle_bound}
% }
\end{align}

In the offering process, TSO compiles a list of selected bids, ordered by price from low to high, until the required quantity is reached to ensure balance. Therefore, the bid prices ($\lambda_{m}^{a,\uparrow}$, $\lambda_{k}^{a,\downarrow}$) for regulating upward and for regulating downward by aggregator $m$ and $k$ are factored into the cost function of TSO.
Besides, $\lambda_{t}^{\uparrow}$ and $\lambda_{t}^{\downarrow}$ are prices of reserve capacity for upward and downward regulation from reserve units. These prices are assumed to be equal to the balancing prices (the imbalance price is derived from the balancing price, and in most cases, when only upward or downward regulation is required, these prices are identical).
The variables $E_{m,t}^{a,\uparrow}$,$E_{k,t}^{a,\downarrow}$ are the volume of flexibility provided by aggregator $m$ and aggregator $k$ at step $t$. 
$E_{t}^{reg,\uparrow}$ and $E_{t}^{reg,\downarrow}$ are the total volume activated for regulation and $E_t^{r}$ is the volume of reserve capacity used. 
$M$ and $K$  are the number of aggregator flexibility units.

Eq. \ref{eq:sum_equal_activated_volume_up} and Eq. \ref{eq:sum_equal_activated_volume_dn} ensure that the sum of the volume provided by the DER flexibility and reserve units is equal to the volume required for upward or downward regulation by the TSO. Equations \ref{eq:up_agg_lim} and \ref{eq:dn_agg_lim} limit the volume activated by the TSO within the boundary that the aggregator can provide.
Equations \ref{eq:agg_sign} and \ref{eq:reserve_sign} represent the sign that the volume regulation for upward regulation is greater than zero and the volume for downward regulation is less than zero.
Eq. \ref{eq:agg_upper_bound}, Eq. \ref{eq:agg_lower_bound}, and Eq. \ref{eq:agg_middle_bound} present the calculation of the boundaries from the output of aggregator models.

% The output of this model is the merit order list of the flexibility volume that will be used from each aggregator and ordered by price from low to high. 
% Figure \ref{fig:fig1} demonstrates a merit order in two different steps where TSO activates upward and downward regulation.

% \begin{figure}[htbp]
% \centerline{\includegraphics[scale=0.52]{images/MOL_example.png}}
% \caption{Example of a figure caption.}
% \label{fig:fig1}
% \end{figure}
\subsection{DSO model}
The DSO will receive the MOL from the TSO or the boundaries from the aggregator. Along with their demand forecast, DSO will perform the power flow analysis to check if any congestion occurs in the network. The DSO's objective function is to minimize the potential cost they have to pay to resolve congestion per ISP, which is shown in Eq. \ref{eq:dso_obj}:
\begin{align}
% \small{
% F^{DSO}_t = 
min\sum_{x\in \Omega_x} (V_{x,t}^\uparrow C_x^\uparrow) -  (V_{x,t}^\downarrow C_x^\downarrow) \label{eq:dso_obj}
% }
\end{align}

In this case, congestion is assumed to arise solely from the activation of downward or upward regulation from the EV aggregators. Consequently, the flexibility deployed to mitigate congestion is drawn from the same aggregation unit as TSO. 
$V_{x,t}^\uparrow, V_{x,t}^\downarrow$ are the volume of flexibility used for managing congestion at node $x$, which is in a set of buses $\Omega_x$. $M$ and $K$ is a subset of $\Omega_x$.
$C_x^\uparrow$ and $C_x^\downarrow$ are congestion prices of flexibility for adjusting upward and downward at node $x$, this price is assumed to be equal to the price for the reserve bid by using the same flexibility unit.
% DSO constraints for each time step in the TSO-DSO hybrid-managed are as follows. 

In the TSO-DSO hybrid-managed model, the DSO uses Eq. \ref{eq:P_up_updated} and Eq. \ref{eq:P_down_updated} to update the demand $P_{x,t}^{d'}$ and generation $P_{x,t}^{g'}$ in the next two ISPs with the volumes used by the TSO from the MOL $E_{x,t}^{a,\uparrow}, E_{x,t}^{a,\downarrow}$. Then, DSO performs the optimal power flow calculation based on Eq. \ref{eq:P_balance}-\ref{eq:line_sus} to determine the flexibility volume $V_{x,t}^\uparrow, V_{x,t}^\downarrow$ to resolve congestion. 
If the optimization calculation yields a feasible solution, this indicates that the available flexibility from the aggregators is sufficient to resolve the congestion caused by the TSO's service.
Otherwise, the boundary for the TSO is divided by $step$-th in \ref{eq:updated_bound_1a} and \ref{eq:updated_bound_1b}. TSO can only use resources within the new constraints. The process is then repeated until the DSO finds the optimal solution or reaches the stopping condition after five divisions, the $E_{m,t}^{a,\uparrow '}, E_{k,t}^{a,\downarrow '}$ are reset to 0.
\begin{align}
% \small{
P_{x,t}^{g'} = P_{x,t}^{g} + E_{x,t}^{a,\uparrow'}  &\quad\forall{x,t} \label{eq:P_up_updated}\\
P_{x,t}^{d'} = P_{x,t}^{d} - E_{x,t}^{a,\downarrow'} &\quad\forall{x,t} \label{eq:P_down_updated}\\
V_{x,t}^\uparrow + V_{x,t}^\downarrow + P_{x,t}^{g'} - P_{x,t}^{d'} = \sum_{y\in J_x} Flow_{xy,t} &\quad\forall{x,y,t} \label{eq:P_balance}\\
\underline{E_{x,t}^{a,\uparrow}} \leq V_{x,t}^{\uparrow} \leq \overline{E_{x,t}^{a,\uparrow}} &\quad\forall {x,t} \label{eq:flex_lim1}\\
\underline{E_{x,t}^{a,\downarrow}} \leq V_{x,t}^{\downarrow} \leq \overline{E_{x,t}^{a,\downarrow}} &\quad\forall {x,t} \label{eq:flex_lim2}\\
V_{x,t}^{\downarrow} \leq 0 \leq V_{x,t}^{\uparrow} &\quad\forall {x,t} \label{eq:flex_lim3}\\
% A^\uparrow_\text{min} \leq A^\uparrow \leq A^\uparrow_\text{max}\label{eq:flex_lim1}\\
% -A^\downarrow_\text{max} \leq A^\downarrow \leq -A^\downarrow_\text{min}\label{eq:flex_lim2}\\
% A^\downarrow < 0, \quad A^\uparrow > 0\label{eq:flex_lim3}\\
Flow_{xy,t} = -Flow_{yx,t} &\quad\forall {x,y,t} \label{eq:p_sym}\\
Flow_{xy,t} < Rated_{xy} PF &\quad\forall {x,y,t} \label{eq:thermal_rating}\\
Flow_{xy,t} = Base_{MVA} B_{xy} (\theta_{x,t} - \theta_{y,t}) &\quad\forall {x,y,t} \label{eq:v_angle}\\
B_{xy} \approx X_{xy} / (R_{xy}^2 + X_{xy}^2) &\quad\forall {x,y} \label{eq:line_sus}
% divide by step size
\end{align}
\begin{align}
E_{x,t}^{a,\uparrow'} = E_{x,t}^{a,\uparrow}/step - V_{x,t}^\uparrow  &\quad\forall{x,t} \label{eq:updated_bound_1a}\\
E_{x,t}^{a,\downarrow'} = E_{x,t}^{a,\downarrow}/step - V_{x,t}^\downarrow  &\quad\forall{x,t} \label{eq:updated_bound_1b}
% new bound
% E_{x,t}^{a,\uparrow} = E_{x,t}^{a,\uparrow} - V_{x,t}^\uparrow  &\quad\forall{x,t} \label{eq:new_bound_1a}\\
% E_{x,t}^{a,\downarrow} = E_{x,t}^{a,\downarrow} - V_{x,t}^\downarrow    &\quad\forall{x,t} \label{eq:new_bound_1b}
% }
\end{align}
% In the DSO-managed coordination scheme, the DSO constraints for each time step are as follows:
% \begin{align}
% % \small{
% % update P d,g in scheme 2
% P_{x,t}^g = P_{x,t}^g + \overline{E_{x,t}^{a,\uparrow'}}   &\quad\forall{x,t} \label{eq:P_up_updated_2}\\
% P_{x,t}^d = P_{x,t}^d - \underline{E_{x,t}^{a,\downarrow'}} &\quad\forall{x,t} \label{eq:P_down_updated_2}\\
% % \overline{E_{x,t}^{a,\uparrow}} = \overline{E_{x,t}^{a,\uparrow}}/step  &\quad\forall{x,t} \label{eq:updated_bound_2a}\\
% % \underline{E_{x,t}^{a,\downarrow}} = \underline{E_{x,t}^{a,\downarrow}}/step  &\quad\forall{x,t} \label{eq:updated_bound_2b}\\
% \overline{E_{x,t}^{a,\uparrow'}} = \overline{E_{x,t}^{a,\uparrow}}/step - V_{x,t}^\uparrow  &\quad\forall{x,t} \label{eq:new_bound_2a}\\
% \underline{E_{x,t}^{a,\downarrow'}} = \underline{E_{x,t}^{a,\downarrow}}/step - V_{x,t}^\downarrow    &\quad\forall{x,t} \label{eq:new_bound_2b}
% % }
% \end{align}

Differently, in the DSO-managed coordination scheme, DSO updates the demand and generation at each node with $\overline{E_{x,t}^{a,\uparrow}},\underline{E_{x,t}^{a,\downarrow}}$ from the aggregator.
% which is presented in Eq. \ref{eq:P_up_updated_2} and Eq. \ref{eq:P_down_updated_2}. 
% The new boundary is obtained by reducing the aggregator boundary with the feasible solution from DSO optimal power flow calculation $V_{x,t}^\uparrow$ and $V_{x,t}^\downarrow$ in Eq. \ref{eq:new_bound_2a} and Eq. \ref{eq:new_bound_2b}, otherwise, this value is gradually reduced with $step^{th}$ iteration.% as in the previous case.
Active power balance condition at node $x$ at step $t$ is presented in Eq. \ref{eq:P_balance}. $J_x$ refers to the subset of nodes that are connected to node $x$. This constraint needs to be satisfied for all nodes in the network.
Similar to the TSO model, the constraints on flexibility boundaries are presented in Eq. \ref{eq:flex_lim1}, Eq.  \ref{eq:flex_lim2}, and Eq. \ref{eq:flex_lim3}.
Power flow symmetry condition in Eq. \ref{eq:p_sym} indicates that the flow of power from node $x$ to node $y$ ($Flow_{xy}$) is opposite to the value of flow from node $y$ to node $x$.
This value is determined by the base power for the network (in MVA) $Base_{MVA}$, the susceptance of line in per unit $B_{xy}$ and the voltage phase angle difference  $\theta_{x,t}$ and $\theta_{y,t}$. 
$B_{xy}$ is calculated by resistance and reactance of the line in Eq. \ref{eq:line_sus}.
Moreover, the rated value of each branch ($Rated_{xy}$) in Eq. \ref{eq:thermal_rating} multiplied by the power factor $PF$ forms the thermal rating threshold for each branch.
The power factor $PF$ is considered 98\% in this study. The
optimization problems are mixed-integer linear programming (MILP) and are solved by the Gurobi solver in Pyomo.

% \begin{figure}[htbp]
% \centerline{\includegraphics[scale=0.8]{images/DSO_priority_flowchart.png}}
% \caption{TSO priority.}
% \label{fig}
% \end{figure}

\section{Case studies}
% - Describe the network, data use (price)
% - Aggregator data
% - Analyse cost
% - Simulation in different days
% - Analyze aggregator behavior
The proposed coordination schemes are validated using the real distribution network at Bleiswijk station in the northern Netherlands. This network comprises 4 step-down two-winding transformers from 25kV to 10.5kV, 182 busbars, 867 terminals, 224 lines, 25 synchronous machines, 170 substations, and 166 loads.
The total installed capacity is 66.06 MW, with a total spinning reserve is 11.36 MW. The PV penetration rate is 12.85\% (8.5MW) from 16 installation points.
The aggregators providing upward regulation service are connected to nodes 12, 42, 145, 146, and 147. While the aggregators providing downward regulation service are connected to nodes 15, 18, 41, 179, and 183 in the network.
% as illustrated in orange and yellow blocks in fig.\ref{fig:fig4}.
Each aggregator manages 100 EVs and offers the same total volume of flexibility. The only difference lies in their bid prices, which are presented in Table I.
% One week in summer, one week in winter, and different BRP prices are simulated to consider the aggregator behavior.
% in price fluctuations and seasonal changes.

\begin{table}[b]
\centering
\caption{The DER flexibility for balancing service}
\fontsize{6.3}{7.5}\selectfont
% \begin{center}
\begin{tabular}{|m{0.5cm}|m{1.2cm}|m{0.6cm}|m{1.2cm}|m{1cm}|}
\hline
\multirow{2}{*}

\textbf{Unit} & 
\textbf{Name} &
\textbf{Node} & 
\textbf{Type} & 
\textbf{Price}
\\
& & & & (\euro/MWh)
\\
\hline
        1 & EV\_Agg1 & 12 & upward & 25 \\ 
\hline
        2 & EV\_Agg2 & 42 & upward & 30 \\ 
\hline
        3 & EV\_Agg3 & 145 & upward & 20 \\  
\hline
        4 & EV\_Agg4 & 146 & upward & 40 \\ 
\hline
        5 & EV\_Agg5 & 147 & upward & 35 \\
\hline
        6 & EV\_Agg6 & 18 & downward & 5 \\ 
\hline
        7 & EV\_Agg7 & 15 & downward & 10 \\ 
\hline
        8 & EV\_Agg8 & 179 & downward & 15 \\  
\hline
        9 & EV\_Agg9 & 41 & downward & -5 \\ 
\hline
        10 & EV\_Agg10 & 183 & downward & -10 \\  
\hline
\end{tabular}
\label{tab1}
% \end{center}
\end{table}

% \begin{table}[htbp]
% \centering
% \caption{The DER flexibility for downward regulation}
% %\fontsize{6.3}{7.2}\selectfont
% % \begin{center}
% \begin{tabular}{|m{0.5cm}|m{1.2cm}|m{0.6cm}|m{1.2cm}|m{1cm}|}
% \hline
% \multirow{2}{*}
% \textbf{Unit} & 
% \textbf{Name} &
% \textbf{Node} & 
% \textbf{Type} & 
% \textbf{Price}
% \\
% & & & & (\euro/MWh)
% \\
% \hline
%         1 & EV\_Agg6 & 18 & downward & 5 \\ 
% \hline
%         2 & EV\_Agg7 & 15 & downward & 10 \\ 
% \hline
%         3 & EV\_Agg8 & 179 & downward & 15 \\  
% \hline
%         4 & EV\_Agg9 & 41 & downward & -5 \\ 
% \hline
%         5 & EV\_Agg10 & 183 & downward & -10 \\  
% \hline
% \end{tabular}
% \label{tab2}
% % \end{center}
% \end{table}
% \begin{figure}[htbp]
% \centerline{\includegraphics[width=1.\columnwidth]{images/single_line_diagram.png}}
% \caption{{Location of aggregator units in the grid.}}
% \label{fig:fig4}
% \end{figure}
\subsection{Aggregator behavior by market price and season}
\begin{figure}[b]
\centerline{\includegraphics[width=1\columnwidth]{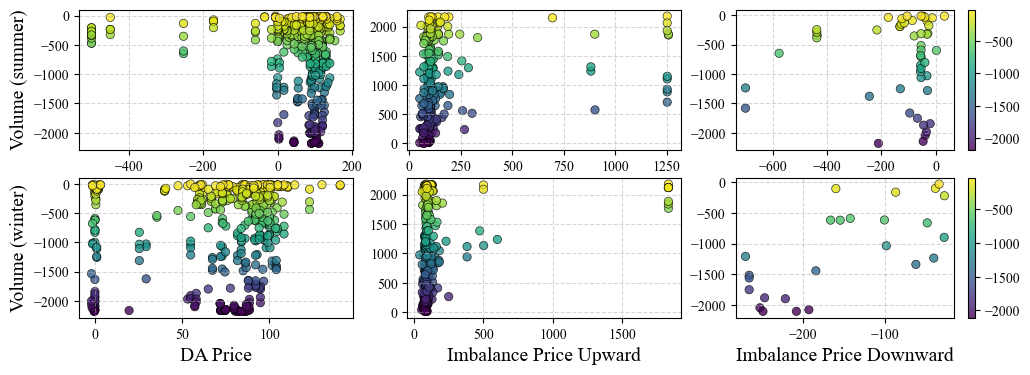}}
\caption{Distribution of volume by price in summer and winter.}
\label{fig:scatter_winter}
\end{figure}

\begin{figure}[b]
\centerline{\includegraphics[width=0.64\columnwidth]{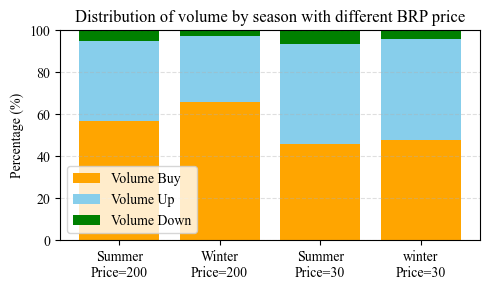}}
\caption{Distribution of volume by season and BRP price.}
\label{fig:pie_chart}
\end{figure}

One week in summer 2023 (June 26th to July 2nd) and one week in winter 2024 (January 1st to January 7th) are selected for the analysis. With a BRP price of 30 \euro/MW and a consumer price of 85 \euro/MW, respectively, which is lower and higher than the average day-ahead price this week.
Both in winter and summer, the occurrence of downward service by downward price is very low and has the lowest total volume, followed by upward service, and the highest is the volume purchased from the DA market, which is presented in Fig. \ref{fig:scatter_winter} and Fig. \ref{fig:pie_chart}. 
% Besides, the downward volume is densely distributed in the region where the upward price is zero and vice versa sparsely distributed in the high upward price region. The same behavior is for the distribution of upward volume versus downward price.
% In most cases in reality, when the system is in deficit, the upward regulation price will increase to attract more volume from service providers, the downward regulation at that time will aggravate the situation of the system. Conversely, there is no need for increased volume when the system is surplus, and the downward regulation price goes very low. 
% \begin{figure}[htbp]
% \centerline{\includegraphics[width=1.\columnwidth]{images/distribution of all volume and price.png}}
% \caption{Distribution of volume activated by prices.}
% \label{fig:dis_vol_all}
% \end{figure}
% The results in Fig. \ref{fig:scatter_summer} and Fig. \ref{fig:scatter_winter} show that in general, the distribution is more concentrated in the occurrence of upward and buy from DA market and less in the downward regulation part. Besides, the fluctuation of downward and day-ahead prices in summer is greater than in winter. 
The distribution of purchase volume by DA price and downward regulation volume by downward price (three figures in the first row) is broader in summer compared to winter (three figures in the second row), primarily due to energy surplus from solar power plants. These surpluses, characterized by zero-marginal prices, lead to many negative price cases for DA and downward price. Contrarily, it's wider for the distribution of upward regulation service by the upward price range caused by the energy shortage in winter. 
% The results show that upward volume concentrates more on a high day-ahead price (100 to 150) region and downward volume is distributed most in the negative price region.
% \begin{figure}[t]
% \centerline{\includegraphics[width=1\columnwidth]{images/scatter_plot_vol_by_price_summer.png}}
% \caption{Distribution of volume by price in summer.}
% \label{fig:scatter_summer}
% \end{figure}
% However, in summer, more upward volume is distributed in high-price regions to get more benefit, while in winter more upward volume is activated but relatively evenly distributed by price due to the system balance being more priority than profit.

\begin{figure}[b]
\centerline{\includegraphics[scale=0.41]{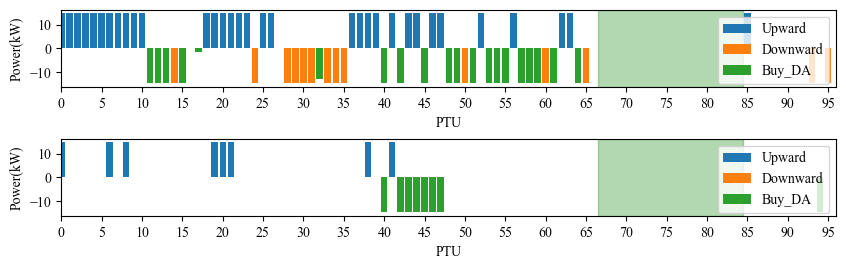}}
\caption{Volume by type of EV1 with BRP price = 30\euro/MWh (on the top) and price = 200\euro/MWh (on the bottom).}
\label{fig:1ev_30}
\end{figure}

In winter, although the price upward regulation is high, the decrease in demand for downward service leads to a decrease in upward service, which is demonstrated in  Fig. \ref{fig:pie_chart}.
Additionally, when BRP prices are high, downward volume (green part) and upward volume (blue part) significantly decrease, while buy volume (orange part) increases, especially in winter.
BRP price impact is also illustrated in the total volume upward, downward regulation, and buy from the DA market of each EV. Fig. \ref{fig:1ev_30} on the top and bottom shows the different behavior by time (PTU) of an individual EV when the BRP price increases from 30 \euro/MWh to 200 \euro/MWh. The orange bars presenting the downward regulation service almost disappear. Additionally, the volume for upward (blue bar) is also decreased, leading to the volume bought from the DA market (green bar) being reduced, because less energy needs to be procured to compensate for the volume discharge when providing upward service.

\subsection{Grid constraint evaluation via two coordination schemes}
\begin{figure}[t]
\centerline{\includegraphics[width=0.92\columnwidth]{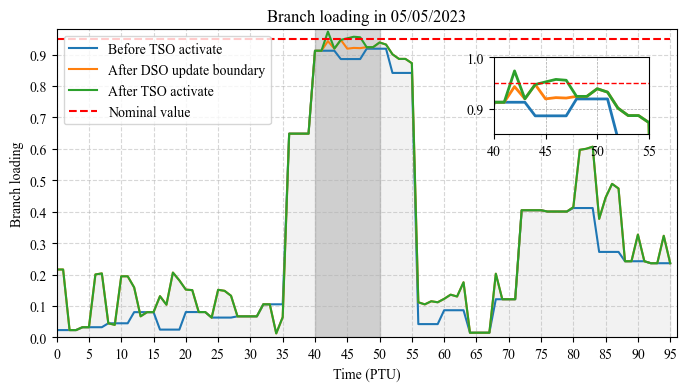}}
\caption{TSO-DSO hybrid-managed.}
\label{fig:tso_priority}
\end{figure}

Providing services brings more benefits to aggregators and prosumers.
However, a problem arises when the TSO activates the service without considering the distribution network constraints, and multiple aggregators react at the same time with very high or very low attractive prices, resulting in congestion within the distribution network. The green line in Fig. \ref{fig:tso_priority} shows that activating the TSO balancing service at ISP 45 causes overloading in the main line straight to the connection point with the external grid. The congestion above is due to the regulation upward from TSO, the line loading is over 95\% due to the upward regulation when the distribution grid is in a low-demand situation, causing a reversed flow from the distribution grid to the transmission grid. This will cause rapid deterioration of the device or damage if the overload condition is prolonged. The dashed red line presents the limitation of the line loading, which is 95\% in this study.
With different coordination between TSO and DSO, these issues are addressed through different sequences.

\begin{figure}[htbp]
  \centering
    \includegraphics[width=0.45\columnwidth]{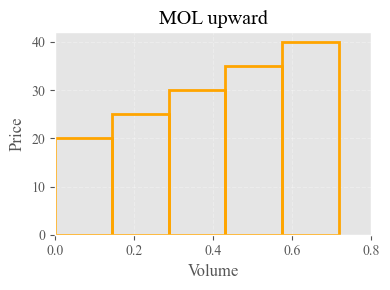}
    \includegraphics[width=0.5\columnwidth]{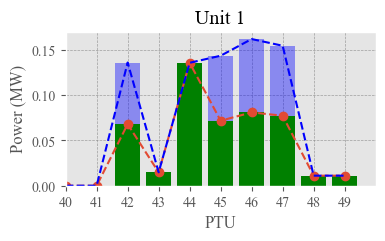}
    \caption{a) MOL upward, b) Flexibility boundaries}
    \label{fig:mol_up}
\end{figure}

In the hybrid scheme, TSO performs an initial filter to select bids before forwarding them to DSO under the MOL form. After DSO validation, the DSO returns updated limits on the volume of flexibility that TSO can use. An example of the MOL at ISP 45 is presented in Fig. \ref{fig:mol_up}a. The flexibility boundaries before and after the validation, from ISP 40 to 50, are depicted in Fig. \ref{fig:mol_up}b by the blue and red lines, respectively. Loading at the observed line after TSO use flexibility with the new boundaries remains below the constraints threshold, which is presented by the orange line in Fig. \ref{fig:tso_priority}.

% \begin{figure}[htbp]
% \centerline{\includegraphics[scale=0.52]{images/MOL_example.png}}
% \caption{Example of a figure caption.}
% \label{fig:MOL}
% \end{figure}

\begin{figure}[t]
\centerline{\includegraphics[width=0.92\columnwidth]{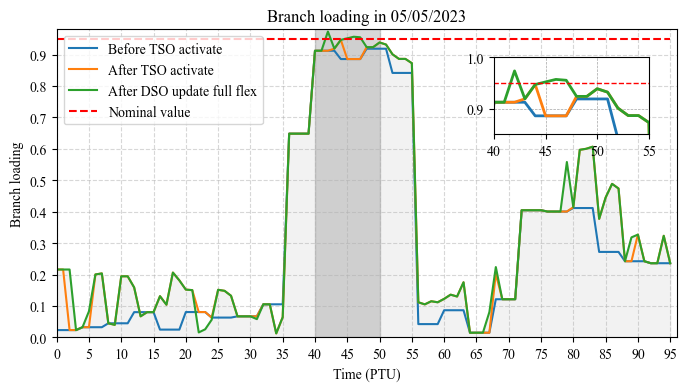}}
\caption{DSO-managed.}
\label{fig:dso_priority}
\end{figure}

Differently, in the DSO-managed model, the aggregator directly sends their flexibility offers to the DSO for validation. DSO performs power flow analysis with the flexibility volume. Then the updated limits are sent to TSO, the simulation results are presented in Fig. \ref{fig:dso_priority}. Similar to the previous case, before the validation, when DSO updates the forecasted demand with the aggregator offers, there is overloading occurs at ISP 45. Subsequently, the DSO performs a linearized AC optimal power flow (ACOPF) to determine the amount of flexibility required to mitigate congestion, iteratively reducing the flexibility boundaries until a feasible solution is obtained. These updated boundaries are then sent to the TSO. The TSO dispatches bids based on system needs while adhering to the new flexibility limits. The resulting loading at the monitored line after TSO activation is shown by the orange line, which confirms that the grid constraint is now satisfied.

However, the difference between the two approaches is not only the order of validation and dispatch actions, but also the amount of flexibility that the aggregator can provide. In the second case, after DSO validation, the loading is lower than in the previous case. This is because the DSO does not know exactly where the TSO will call services, reduces the boundary across all flexibility units until the congestion can be resolved by the available flexibility. This approach leads to the loss of profit for aggregators if more reserve energy can not be used.

A cost analysis was conducted for both the TSO and the aggregator across different days, are highlighted in Table II. Results show that the hybrid model offers greater benefits for both TSO and the aggregator. Since the DER offer price is cheap, the more reserve energy is used from DER flexibility, the lower the cost to the TSO. Therefore, the hybrid model helps TSO reduce costs by accessing a larger volume of DER flexibility than the DSO-managed model. Furthermore, while more flexibility is provided, the aggregator earns more profit, which explains the higher benefit in the hybrid model compared with the DSO-managed model. Although the reduction in TSO costs may not be significant, the increase in aggregate benefits, ranging from 2\% to 11\%, is worth considered.
Besides the economic benefits, the hybrid scheme requires more complex decision process between TSO and DSO, which may lead to lower DER dispatch efficiency. Furthermore, more information needs to be shared between TSO and DSO, which might require more investment in cybersecurity management.

\begin{table}[b]
\fontsize{7}{7.2}\selectfont
\centering
\caption{TSO Cost and Aggregator Benefit under Different coordination schemes}
\begin{tabular}{|c|cc|cc|}
\hline
% \rowcolor{cyan!20}
\textbf{} & \multicolumn{2}{c|}{\textbf{TSO cost (\euro)}} & \multicolumn{2}{c|}{\textbf{Aggregator benefit (\euro)}} \\
% \rowcolor{cyan!20}
\textbf{Date} & \textbf{Hybrid} & \textbf{DSO-managed} & \textbf{Hybrid} & \textbf{DSO-managed} \\
\hline
05/May & 1.353.194 & 1.354.360 ($\uparrow0.1\%$) & 15.953 & 14.786 ($\downarrow7.3\%$)\\
18/Jul & 671.354  & 671.681 ($\uparrow0.1\%$)  & 17.614 & 17.284 ($\downarrow1.9\%$)\\
14/Aug & 68.642   & 68.908 ($\uparrow0.4\%$)  & 2.614  & 2.349 ($\downarrow10.1\%$) \\
05/Sep & 690.873  & 692.622 ($\uparrow0.3\%$) & 15.942 & 14.187 ($\downarrow11.0\%$) \\
06/Sep & 614.466  & 615.237 ($\uparrow0.1\%$) & 19.435 & 18.661 ($\downarrow4.0\%$)\\
\hline
\end{tabular}
\end{table}

\section{Conclusion}
To effectively harness the potential of the DERs' services in the current context, coordination between TSOs and DSOs is necessary to maximize social welfare as well as minimize the risks of conflicting use cases. Notably, DSOs play a more important role in validating services before being utilized by TSOs. This validation ensures that grid constraints are taken into account when providing services from aggregators under the DSO's management. Thereby, mitigating the potential risks of unexpected overload caused by the simultaneous reaction of aggregators with prices and requests from TSOs.
% correct this sentence
% Analyzing the service provision from multiple aggregators shows that, in the case where aggregators have complete information on prices and EV sessions, they will try to optimize the bidding strategies to maximize their profits. Similar behavior from multiple aggregators in close region in response to attractive very high upward regulation price or very low downward regulation price will lead to peak revert flow from their connection point to the grid with higher voltage level or vise versa. 
% The aggregator's objective is to maximize their profits by optimizing bidding strategies. 
Analyzing the service provision from the aggregator shows that similar behavior from multiple aggregators in neighboring regions in response to the attractive very high upward regulation price or very low downward regulation price will lead to excessive reversed power flow from their connection points to the higher voltage grid or vice versa.
%%%
Besides, the price the aggregator contracts with BRP for their real-time deviation has a great impact on the decision to provide services and the final profit. 
The reduction of downward service demand in winter results in the reduction of upward service and volume purchase from the DA market to compensate for the discharge volume. The coordination between TSO and DSO in the offering process allows the provision of services within the grid constraints. 
% correct this sentence
% When TSO has the priority of receiving information, they create the MOL and DSO will update the boundary from the selected bid, this lead to more volume from EVs aggregator is used. On the contrary, if DSO receive information of the flexibility first, they gradually decrease the boundary of all units until no violation occurs, this lead to the greater amount of service is excluded. 
% In the TSO-DSO hybrid-managed model, the DSO updates the boundaries based on the TSO's selected bids. Conversely, in the DSO-managed model, unused bids are considered when DSOs resolve congestion, resulting in a larger portion of services being excluded in this scenario compared with the first one.
Within the scope of this study, the cost-benefit analysis shows that the hybrid-managed scheme gives an advantage with more volume provided and more benefit to the aggregator and lower cost to the TSO than the DSO-managed scheme. 
%%%
The risks associated with inaccurate forecasts of demand, price, and EV sessions significantly influence the decisions of each participant will be addressed in future studies.

% \section*{Acknowledgment}
% The authors would like to acknowledge the financial support for this work from the Netherlands Organization for Scientific Research (NWO) funded DEMOSES project.
% \bibliographystyle{plain}
\bibliography{references}
\end{document}